\documentclass{article}
\usepackage{graphicx}

\input{tcilatex}
\begin{document}

\textbf{Geometrical representation of real and reactive powers of load
demand by orbit diagrams in the Mandelbrot set.}

H\'{e}ctor A. Tabares-Ospina\footnote{%
Facultad de Ingenier\'{\i}a, Instituci\'{o}n Universitaria Pascual Bravo,
Medell\'{\i}n, Colombia.
\par
Email: h.tabares@pascualbravo.edu.co} and John E. Candelo-Becerra\footnote{%
Facultad de Minas, Universidad Nacional de Colombia, Medell\'{\i}n, Colombia.
\par
Email: jecandelob@unal.edu.co}\vspace{0.5cm}

\bigskip \textbf{Abstract}

This paper presents the geometrical representation of the load demand by
using orbits diagrams in the Mandelbrot set, to identify changing behaviors
during a day period of the real and reactive powers. To perform this,
different power combinations were used to represent the fractal diagrams
with an algorithm that considers the mathematical model of Mandelbrot set
and orbits diagrams. A qualitative analysis of the orbits is performed to
identify the fractal graphic patterns with respect to the real and reactive
power consumptions. The results show repetitive graphic patterns in the
fractal space of the power consumption during the day, which help represent
the consumption behavior on a daily load demand curve. The orbit diagrams
save form and structure relations during the daily behavior of the power
consumption. This work shows a different method of evaluating load demand
behavior by using orbit diagrams as a potential tool that will lead to
identify load behavior, useful in operational decisions and power system
planning.


\section{Introduction}

Mathematician Benoit Mandelbrot has defined the concept of fractals as a
semi-geometric element with a repetitive structure at different scales [1],
with characteristics of self-similarity as seen in some natural formations
such as snowflakes, ferns, peacock feathers, and romanesco broccoli. Fractal
theory has been applied to various fields such as biology [2,3], health
sciences [4--8], stock markets [9], network communications [10--12], and
others. Fractal theory is one of the methods used to analyze data and obtain
relevant information in highly complex problems. Thus, it has been used to
study the price of highly variable markets, which are not always explainable
from classical economic analysis.

For example, in [9], the authors demonstrate that current techniques have
some issues to explain the real market operation and a better understanding
is achieved by using techniques such as chaos theory and fractals. In their
publication, the authors show how to apply fractal behavior to stock markets
and refer to multifractal analysis and multifractal topology. The first
describes the invariability of scaling properties of time series and the
second is a function of the H\"{o}lder exponents that characterizes the
degree of irregularity of the signal, and their most significant parameters.

In [13], the authors discuss the basic principle of fractal theory and how
to use it to forecast the short-term electricity price. In the first
instance, the authors analyze the fractal characteristic of the electricity
price, confirming that price data have this property. In the second
instance, a fractal model is used to build a forecasting model, which offers
a wide application in determining the price of electricity in the markets.

Similarly, the authors of [14] demonstrate that the price of thermal coal
has multifractal features by using the concepts introduced by
Mandelbrot-Bouchaud. Hence, a quarterly fluctuation index (QFI) for thermal
power coal price is proposed to forecast the coal price caused by market
fluctuation. This study also provides a useful reference to understand the
multifractal fluctuation characteristics in other energy prices.

Fractal geometry analysis has been also applied to study the morphology and
population growth of cities, and to electricity demand related to the
demography of cities. In [15], a multifractal analysis is used to forecast
electricity demand, explaining that two fractals are found that reflect the
behavior pattern of load demand. Two concepts linked to fractal geometry are
fractal interpolation and extrapolation, which are related to the resolution
of a fractal-encoded image. In [16], an algorithm is used to forecast the
electric charge in which fractal interpolation and extrapolation are also
involved; for the forecast dataset, the average relative errors are only
2.303\% and 2.296\%, respectively, indicating that the algorithm has
advantages in improving forecast accuracy.

In [17] a design method of antenna array consisting of eight microstrip
patches modified with Sierpinski fractal curves has been presented and
experimentally validated in this paper. Method proposed has enabled the
achievement of considerable miniaturization of array length (26\%), together
with multi-band behavior of the antenna, which proves the attractiveness of
presented design methodology and its ability to be implemented in more
complex microstrip structures.

In [18] is studied the application of Triangular Prism Method (TPM)
algorithm in computer assisted Papanicolaou smears analysis that is useful
in cervical cancer screening. The TPM algorithm allows estimation of the FD
(fractal dimension) for optical density of cell nuclei. Selection of the
local FD for green color channel gives efficient separation between both
cell nuclei classes. Proposed algorithm (Tiled TPM) improves separation by
the fractal based estimation using larger area of the cell nuclei.

No paper in the extant literature has examined the orbit diagrams in the
Mandelbrot set to represent the daily load demand. Therefore, this work
focuses on studying the behavior of the different combination of complex
numbers corresponding to a daily load demand, which form different orbit
diagrams in the Mandelbrot set. Particularly, in this work, the real and
reactive powers are used to represent the different orbit diagrams
calculated with the Mandelbrot algorithm, perform observations, and analyze
qualitatively fractal geometry related to load demand.

For this reason, this paper proposes the real and reactive powers of load
demand curve can be characterized by orbit diagrams in the Mandelbrot set.
The test focuses on identifying a clear pattern with similarities in the
daily load demand. Besides, this method is proposed to obtain a new way of
visualizing the behavior of load demand curves, identify the stability of
the orbit diagram according to the variation in the electric power
consumption, and identify easily loadability of the power system. The rest
of this document is organized as follows. Section 2 includes a brief
explanation of the theory of the Mandelbrot sets and orbit diagrams. Section
3 presents the results and discusses the most relevant examples of orbits
created for the real and reactive powers. Finally, the main conclusions of
this research work are summarized.

\section{Research method}

An algorithm that creates fractal diagrams applied to the typical load
demand curve is presented, with the aim of identifying patterns from orbit
diagrams in the Mandelbrot set that represent power consumption. Below, this
section shows the general procedure and the algorithms implemented to obtain
the fractals.

\subsection{General procedure}

Figure \ref{Fig1} presents a step-by-step procedure applied to graph the
fractal diagrams from the load demand with the Mandelbrot and Julia sets.
This figure shows that the first step ($P1$) is to convert the initial data
to manage the procedure to the Mandelbrot and Julia algorithms. Next, the
Mandelbrot algorithm is programmed according to the mathematical theory ($P2$%
) to generate a new data set. Besides, the Julia algorithm is also
programmed to perform the generation of the new sets, based on the
Mandelbrot set ($P3$). With these data sets, it is possible to plot the
different fractals ($P4$) which are then analysed to present the different
results in this paper ($P5$) and the corresponding conclusions.

\includegraphics{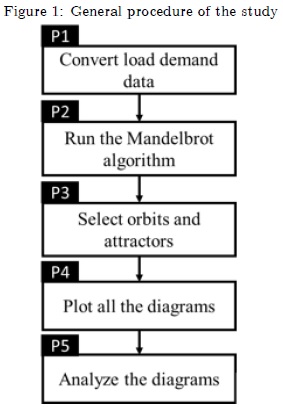}

\bigskip 
\begin{figure*}[h]\centering%
\caption{General procedure of the study}%
\label{Fig1}%
\begin{tabular}{l}
\end{tabular}%
\end{figure*}%

\subsection{Load demand curve}

The load demand has a strong daily pattern that represents the working days
as the data have a very similar demand profile. Thus, the time series is
seasonal because it has a regular repetition pattern during the same period
of time. The periodic behavior is reflected in parameters such as mean,
standard deviation, asymmetry, and autocorrelation asymmetry.

With the purpose to study the orbit diagram of the points related to the
evolution function $Z_{t+1}=Z_{t}^{2}+C$, that moves on the set of
Mandelbrot, according to the law dictated by the load demand curve. The
process begins by reading the typical load demand records of real and
reactive powers defined for a $24$-hour period (see Table \ref{tbl1}). The
per unit values of the load demand are calculated with the following
expression: $Per\_unit\_value=Actual\_MVA/Base\_MVA$. In this case, the base
power is $4000$ MVA. These data are used to plot the diagrams with the
algorithms, in which the lowest and highest consumption points are
considered to evaluate the different fractal diagrams.

\bigskip 
\begin{table*}[ptb]\centering%
\caption{Daily power demand}%
\label{tbl1}%
\begin{tabular}{lllll}
$Hour$ & $P$ & $Q$ & $P_{pu}$ & $Q_{pu}$ \\ 
00:00:00 & 889 & 371 & 0.222 & 0.092 \\ 
01:00:00 & 834 & 405 & 0.208 & 0.101 \\ 
02:00:00 & 792 & 337 & 0.197 & 0.082 \\ 
03:00:00 & 790 & 324 & 0.199 & 0.081 \\ 
04:00:00 & 804 & 323 & 0.201 & 0.080 \\ 
05:00:00 & 925 & 355 & 0.231 & 0.088 \\ 
06:00:00 & 1041 & 482 & 0.260 & 0.120 \\ 
07:00:00 & 1105 & 556 & 0.276 & 0.139 \\ 
08:00:00 & 1191 & 610 & 0.297 & 0.152 \\ 
09:00:00 & 1256 & 704 & 0.314 & 0.176 \\ 
10:00:00 & 1309 & 744 & 0.327 & 0.186 \\ 
11:00:00 & 1366 & 775 & 0.341 & 0.193 \\ 
12:00:00 & 1385 & 793 & 0.346 & 0.198 \\ 
13:00:00 & 1356 & 774 & 0.339 & 0.193 \\ 
14:00:00 & 1337 & 759 & 0.334 & 0.189 \\ 
15:00:00 & 1350 & 774 & 0.337 & 0.193 \\ 
16:00:00 & 1336 & 773 & 0.334 & 0.193 \\ 
17:00:00 & 1312 & 749 & 0.328 & 0.187 \\ 
18:00:00 & 1287 & 687 & 0.321 & 0.171 \\ 
19:00:00 & 1420 & 683 & 0.355 & 0.170 \\ 
20:00:00 & 1389 & 660 & 0.351 & 0.167 \\ 
21:00:00 & 1311 & 605 & 0.327 & 0.151 \\ 
22:00:00 & 1175 & 544 & 0.293 & 0.136 \\ 
23:00:00 & 1030 & 489 & 0.257 & 0.122%
\end{tabular}%
\end{table*}%

\newpage

\subsection{\protect\bigskip Algorithm to create the Mandelbrot set}

Mandelbrot set, denoted as $M=\{$\ $c\in C/J_{c}\}$, represents the sets of
complex numbers $C$ obtained after iterating the from the initial point $%
Z_{n}$ and the selected constant $C$ as shown (\ref{Ec0}), the results form
a diagram with connected points remaining bounded in an absolute value. One
property of $M$ is that the points are connected, although in some zones of
the diagram it seems that the set is fragmented. The iteration of the
function generates a set of numbers called orbits. The results of the
iteration of those points out of the boundary set tend to infinity.

\begin{equation}
Z_{n+1}=F\left( Z_{n}\right) =Z_{n}^{2}+C  \label{Ec0}
\end{equation}

From the term $C$, a successive recursion is performed with $Z_{0}=0$ as the
initial term. If this successive recursion is dimensioned, then the term $C$
belongs to the Mandelbrot set; if not, then they are excluded. Therefore,
Figure \ref{Fig2} shows the Mandelbrot set with points in the black zone
that are called the prisoners, while the points in other colors are the
escapists and they represent the velocity that they escape to infinite.

\includegraphics{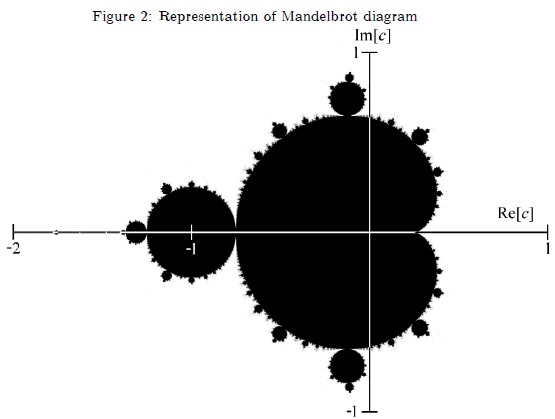}

\begin{figure*}[h]\centering%
\caption{Representation of Mandelbrot diagram}%
\label{Fig2}%
\begin{tabular}{l}
\end{tabular}%
\end{figure*}%

\bigskip From this figure, the number -1 is inside of the set while the
number 1 is outside of the set. In the Mandelbrot set, the fractal is the
border and the dimension of Hausdorff is unknown. If the image is enlarged
near the edge of the set, many areas the Mandelbrot set are represented in
the same form. Besides, different types of Julia sets are distributed in
different regions of the Mandelbrot set. Whether a complex number appears
with a greater value than 2 in the 0 orbit, then the orbit tends to infinity.

The pseudocode of the algorithm that is used to represent the Mandelbrot set
is presented as follows:

\textbf{Start}

\hspace{1cm}For each point $C$ in the complex plane do:

\hspace{1cm}Fix $Z_{0}=0$

\hspace{1cm}\textbf{For} $t=1$ to $t_{\max }$ \textbf{do}:

\hspace{1cm}\hspace{1cm}Calculate $Z_{t}=Z_{t}^{2}+C$

\hspace{1cm}\hspace{1cm}\textbf{If} $|Z_{t}|>2$ \textbf{then}

\hspace{1cm}\hspace{1cm}\hspace{1cm}Break

\hspace{1cm}\hspace{1cm}\textbf{End if}

\hspace{1cm}\hspace{1cm}\textbf{If} $t<t_{\max }$ \textbf{then}

\hspace{1cm}\hspace{1cm}\hspace{1cm}Draw $C$ in white (the point does not
belong to the set)

\hspace{1cm}\hspace{1cm}\textbf{Else if} $t=t_{\max }$ \textbf{then}

\hspace{1cm}\hspace{1cm}\hspace{1cm}Draw $C$ in black (as the point does
belong to the set)

\hspace{1cm}\hspace{1cm}\textbf{End if}

\hspace{1cm}\textbf{End For}

\textbf{End\vspace{0.5cm}}

In this research, the presented algorithm has been used to obtain the
Mandelbrot set and the diagram that represent it. Some points related to the
real and reactive powers with the respective signs are studied in the
Mandelbrot set and related to those points created for the orbits diagrams
as explained in the following sections.

\subsection{Orbit diagrams and attractors}

One way to visualize the state of a system is through the orbit diagram. An
orbit is a set points related with the evaluation function of a dynamic
system. For discrete dynamics systems the orbits are successions. The basic
classifications can be defined as: (a) fix points (b) periodic orbits, and
(c) orbits no constants [19]. With respect to the attractor, the main
properties are a) compression, b) expansion, and c) folding [20].

If there is an attractor in the complex plane, the orbit associated with a
complex number of the form $z=a+bi$ is an orbit of complex numbers, with the
same dynamics. The orbits are a sequence of complex numbers and their
characteristics depend fundamentally on the values of the initial point $%
Z_{n}$ from which it is split and the selected constant $C$. The pseudocode
used to generate orbit diagrams and find the attractor of a complex number
inside the $M$ set is described as follows.

\textbf{Start}

\hspace{1cm}Read $C$

\hspace{1cm}Fix $Z_{0}=C$

\hspace{1cm}\textbf{For} $t=1$ to $t_{MaxNumOrbits}$ \textbf{do}:

\hspace{1cm}\hspace{1cm}Calculate $Z_{t}=Z_{t}^{2}+C$

\hspace{1cm}\hspace{1cm}\textbf{If} $|Z_{t}|>2$ \textbf{then}

\hspace{1cm}\hspace{1cm}\hspace{1cm}Break

\hspace{1cm}\hspace{1cm}\textbf{End if}

\hspace{1cm}\hspace{1cm}Draw orbits of $Z_{t}$

\qquad \qquad \qquad $Z_{t}=Z_{t+1}^{{}}$\hspace{1cm}\hspace{1cm}

\hspace{1cm}\textbf{End For}

\textbf{End\vspace{0.5cm}}

For each hour the number of orbits and the attractor value is shown in the
table \ref{tbl2}.

\bigskip 
\begin{table*}[ptb]\centering%
\caption{Daily power demand, attractors and orbits}%
\label{tbl2}%
\begin{tabular}{lll}
$Hour$ & $Attractor$ & $Num.orbits$ \\ 
00:00:00 & 0.319 & 5 \\ 
01:00:00 & 0.294 & 5 \\ 
02:00:00 & 0.274 & 5 \\ 
03:00:00 & 0.276 & 5 \\ 
04:00:00 & 0.280 & 3 \\ 
05:00:00 & 0.305 & 5 \\ 
06:00:00 & 0.371 & 9 \\ 
07:00:00 & 0.393 & 9 \\ 
08:00:00 & 0.415 & 18 \\ 
09:00:00 & 0.433 & 30 \\ 
10:00:00 & 0.448 & 32 \\ 
11:00:00 & 0.463 & 50 \\ 
12:00:00 & 0.468 & 53 \\ 
13:00:00 & 0.462 & 44 \\ 
14:00:00 & 0.457 & 38 \\ 
15:00:00 & 0.464 & 41 \\ 
16:00:00 & 0.456 & 41 \\ 
17:00:00 & 0.451 & 41 \\ 
18:00:00 & 0.445 & 41 \\ 
19:00:00 & 0.486 & 89 \\ 
20:00:00 & 0.482 & 89 \\ 
21:00:00 & 0.449 & 41 \\ 
22:00:00 & 0.416 & 18 \\ 
23:00:00 & 0.365 & 14%
\end{tabular}%
\end{table*}%

\subsection{Algorithm to study the orbits of the load demand}

In order to obtain the results of the fractal topology patterns that
represent the real and reactive power of the load demand curve, the
procedure shown in Fig. \ref{Fig3} was followed. The algorithm begins by
reading the data of the real and reactive power, in which the comparative
curve can be obtained for the different graphs to be made. Then, with these
same power values and the reading of the power base, the calculation of the
values per unit of power can be made.

A third step corresponds to calculate the $M$ set using the Mandelbrot
algorithm and with this the values of real and reactive powers are adjusted
within the $M$ set. The fourth step was to graph the respective orbit
diagrams associated to the Mandelbrot set. Finally, the fifth step considers
the evaluation of the orbit diagrams related to the daily power consumption.

\includegraphics{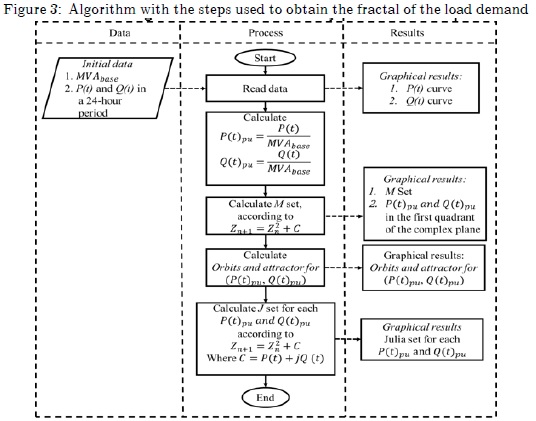}

\bigskip 
\begin{figure*}[h]\centering%
\caption{Algorithm with the steps used to obtain the fractal of the load demand}%
\label{Fig3}%
\begin{tabular}{l}
\end{tabular}%
\end{figure*}%

\bigskip

\section{Results and analysis}

Figure \ref{Fig4} presents the typical load demand curve plotted with the
data of Table 1 and Fig. \ref{Fig5} presents the load demand plotted in the
first quadrant of the complex plane. As real and reactive powers are
positive, they represent a load consumption related to inductive elements.
Under these conditions, the three most interesting values of the power
consumption that are selected are the lowest consumption at $3:00$, the
highest consumption at $19:00$, and the approximate average consumption at $%
09:00$. Other hours of the day represent diagrams that are forms between the
values as shown in the following results in this section.

\includegraphics{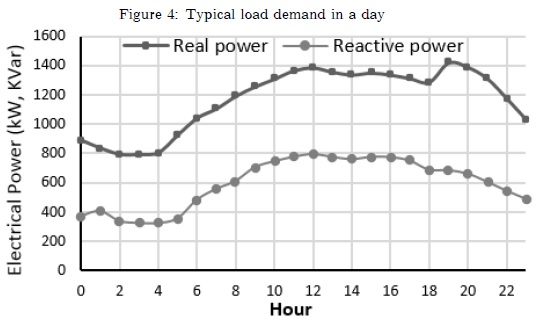}

\bigskip 
\begin{figure*}[h]\centering%
\caption{Typical load demand in a day}%
\label{Fig4}%
\begin{tabular}{l}
\end{tabular}%
\end{figure*}%

Figure \ref{Fig5} shows a very low power consumption value at $03:00$, which
then increases as the hours pass in the day.

\includegraphics{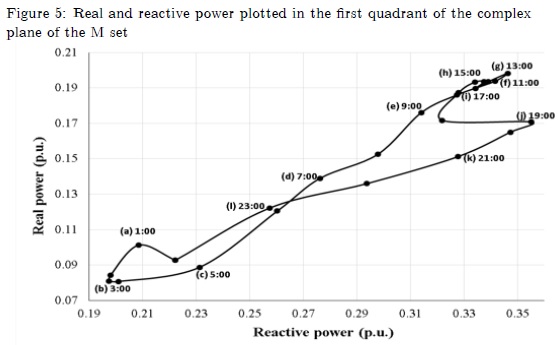}

\begin{figure*}[h]\centering%
\caption{Real and reactive power plotted in the first quadrant of the complex plane of the M set}%
\label{Fig5}%
\begin{tabular}{l}
\end{tabular}%
\end{figure*}%

In addition, it is observed that the curve presents changes during the
different hours of the day, due to the combinations of real and reactive
powers. In some cases, the curves cross indicating that the complex numbers
formed are equal at different times of the day. Due to the number of curves
that can be created with orbit diagrams, only those that are marked in this
curve have been created to obtain an analysis of them.

Figure \ref{Fig6} shows the orbit diagram generated for each point in Fig. %
\ref{Fig5}. These orbits are created by performing iterations of the complex
numbers obtained from the daily load demand (see table \ref{tbl2}).

\includegraphics{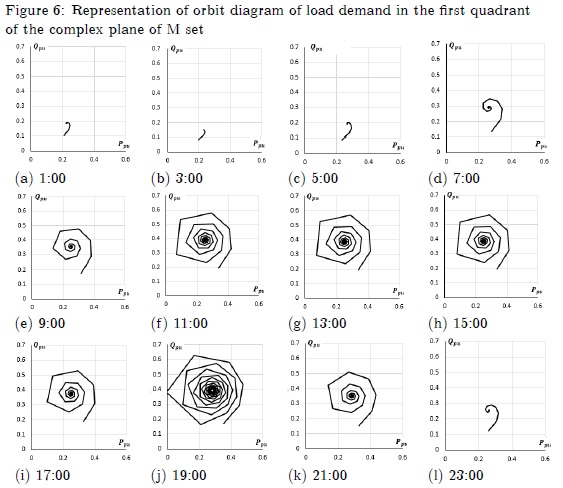}

\begin{figure*}[h]\centering%
\caption{Representation of orbit diagram of load demand in the first quadrant of the complex plane of M set}%
\label{Fig6}%
\begin{tabular}{llll}
&  &  &  \\ 
(a) 1:00 & (b) 3:00 & (c) 5:00 & (d) 7:00 \\ 
&  &  &  \\ 
(e) 9:00 & (f) 11:00 & (g) 13:00 & (h) 15:00 \\ 
&  &  &  \\ 
(i) 17:00 & (j) 19:00 & (k) 21:00 & (l) 23:00%
\end{tabular}%
\end{figure*}%

The process of generating the orbit diagram reveals folds with the following
properties: from $1:00$ to $11:00$ (Fig. \ref{Fig6}a - Fig. \ref{Fig6}f) the
orbits expand causing their space to stretch, associated with the
progressive increase in the hourly load demand. From $11:00$ to $15:00$
(Fig. \ref{Fig6}f -- Fig. \ref{Fig6}h) the orbits remain practically
unaltered, corresponding with a constant demand of the hourly electrical
power. Then, at 17:00 (Fig. \ref{Fig6}i) the orbit is reduced because of the
power reduction and after that at $19:00$ the maximum consumption with a
large orbit is presented (Fig. \ref{Fig6}j). From $21:00$ to $23:00$ (Fig. %
\ref{Fig6}k -- Fig. \ref{Fig6}l) the load demand decreases and the orbit
diagram also reduces.

Fig. \ref{Fig7} shows the number of points obtained with the orbit diagram
algorithm after iterating the complex numbers of real and reactive powers.
The orbit diagram in the complex plane of the Mandelbrot set obeys the daily
periodic curve profile of the load demand, which is continuously reduced
from $20:00$ to $5:00$. In the period $00:00-05:00$, the lower load demand
is presented, which corresponds to orbit diagrams of 5 points without
folding. From $05:00$, the load demand begins to increase and the points are
greater than $10$, expanding the orbits in the complex plane. Besides, the
number of points with highest values correspond to the same peaks in the
load demand curve at $12:00$ and $19:00$. A value close to the peak is
presented at $20:00$, because the consumption remains also in a high value
close to that of the peak at $19:00$.

This figure shows some clear relations between the power consumption and the
number of points created in the orbits. This result helps to understand the
system loadability and the increasing number of points when the system is
reaching the consumption limits, as the expansions of orbits are presented
in the complex plane.

\includegraphics{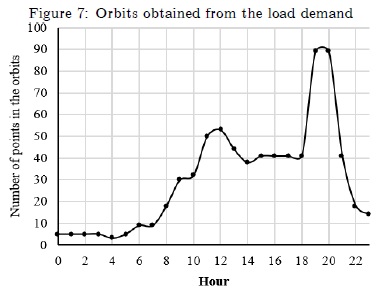}

\begin{figure*}[h]\centering%
\caption{Orbits obtained from the load demand}%
\label{Fig7}%
\begin{tabular}{l}
\end{tabular}%
\end{figure*}%

\newpage

Figure \ref{Fig8} shows the values of the attractor for the different points
in the load demand curve (see table \ref{tbl2}). This figure shows that the
attractor moves according to the behavior of the power consumption, which
help to identify a clear pattern that repeat during different days in the
power consumption. This result also shows that the real and reactive power
consumption can be identified based on the position of the attractor, as the
method helps to identify the maximum values reached by the power load.

\includegraphics{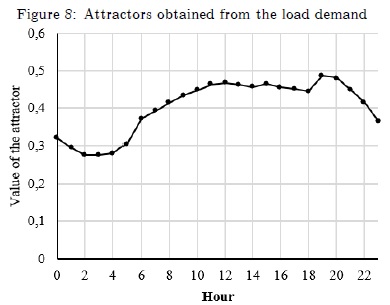}

\begin{figure*}[h]\centering%
\caption{Attractors obtained from the load demand}%
\label{Fig8}%
\begin{tabular}{l}
\end{tabular}%
\end{figure*}%

\newpage

\section{Conclusions}

This paper presented the geometrical representation of the load demand by
using orbits diagrams in the Mandelbrot set, to identify changing behaviors
during a daily load demand. Power combinations were used to represent the
orbit diagrams with an algorithm that considers the mathematical model of
Mandelbrot set and orbit diagrams. The results allowed to identify a new
space of analysis of the discrete dynamic system based on the daily load
demand of real and reactive powers, by means of orbit diagrams in the
Mandelbrot set. The orbit diagrams in the complex plane of the Mandelbrot
set obtained with real and reactive powers of the load demand curve
established the behavior of the iterated sequence for given values of $%
(P_{0},Q_{0})$, interpreting the orbit geometry obtained, allowing to
quickly discern the behavior of the load demand. As the load demand changes
constantly in the daily load demand, the orbit diagrams create small number
of points because of the small power consumption and create large number of
points because of large power consumption. Stability of the periodic orbits
appears and disappears with the power variation in the daily demand curve.
The density of the folding of the orbits is related to the proximity of the
load demand of real and reactive powers to the limits of the Mandelbrot set.

\section{Acknowledgement}

This work was supported by the Agencia de Educaci\'{o}n Superior de Medell%
\'{\i}n (Sapiencia), under the specific agreement celebrated with the
Instituci\'{o}n Universitaria Pascual Bravo. The project is part of the
Energy System Doctorate Program and the Department of Electrical Energy and
Automation of the Universidad Nacional de Colombia, Sede Medell\'{\i}n,
Facultad de Minas.

\section{References}

[1]\qquad Barnsley M.F., Devaney R.L., Mandelbrot B.B., Peitgen H.-O., Saupe
D., Voss R.F., The Science of Fractal Images, New York, NY: Springer New
York; (1988). doi:10.1007/978-1-4612-3784-6.

[2]\qquad Strogatz S.H., Nonlinear Dynamics and Chaos. CRC Press, (2018).
doi:10.1201/9780429492563.

[3]\qquad Losa G.A., Fractals and Their Contribution to Biology and
Medicine. Medicographia, 34(2012), 365--374.

[4]\qquad Garcia T.A., Tamura Ozaki G.A., Castoldi R.C., Koike T.E.,
Trindade Camargo R.C., Silva Camargo Filho J.C., Fractal dimension in the
evaluation of different treatments of muscular injury in rats, Tissue Cell,
54(2018),120--6. doi:10.1016/j.tice.2018.08.014.

[5]\qquad Rodr\'{\i}guez V. J.O., Prieto B. S.E., Correa H. S.C., Soracipa
M. M.Y., Mendez P. L.R., Bernal C. H.J., et al., Nueva metodolog\'{\i}a de
evaluaci\'{o}n del Holter basada en los sistemas din\'{a}micos y la geometr%
\'{\i}a fractal: confirmaci\'{o}n de su aplicabilidad a nivel cl\'{\i}nico,
Rev La Univ Ind Santander Salud, 48(2016), 27--36.
doi:10.18273/revsal.v48n1-2016003.

[6]\qquad Popovic N., Radunovic M., Badnjar J., Popovic T., Fractal
dimension and lacunarity analysis of retinal microvascular morphology in
hypertension and diabetes, Microvasc Res, 118(2018), 36--43.
doi:10.1016/j.mvr.2018.02.006.

[7]\qquad Hern\'{a}ndez Vel\'{a}zquez J. de D., Mej\'{\i}a-Rosales S., Gama
Goicochea A., Fractal properties of biophysical models of pericellular
brushes can be used to differentiate between cancerous and normal cervical
epithelial cells. Colloids Surfaces B Biointerfaces, 170(2018), 572--577.
doi:10.1016/j.colsurfb.2018.06.059.

[8]\qquad Moon P., Muday J., Raynor S., Schirillo J., Boydston C., Fairbanks
M.S., et al., Fractal images induce fractal pupil dilations and
constrictions. Int J Psychophysiol, 93(2014), 316--21.
doi:10.1016/j.ijpsycho.2014.06.013.

[9]\qquad Mandelbrot BB, Hudson RL. The (mis) Behaviour of Markets: A
Fractal View of Risk, Ruin and Reward. London: Profile Books, (2004).

[10]\qquad Kumar R., Chaubey P.N., On the design of tree-type ultra wideband
fractal Antenna for DS-CDMA system. J Microwaves, Optoelectron Electromagn
Appl, 11(2012),107--21. doi:10.1590/S2179-10742012000100009.

[11]\qquad Ma Y-J., Zhai M-Y., Fractal and multi-fractal features of the
broadband power line communication signals, Comput Electr Eng, (2018)
doi:10.1016/j.compeleceng.2018.01.025.

[12]\qquad Ye D., Dai M., Sun Y., Su W., Average weighted receiving time on
the non-homogeneous double-weighted fractal networks, Phys A Stat Mech Its
Appl, 473(2017), 390--402. doi:10.1016/j.physa.2017.01.013.

[13]\qquad Cui H., Yang L., Short-Term Electricity Price Forecast Based on
Improved Fractal Theory, IEEE Int. Conf. Comput. Eng. Technol., (2009),
347--51. doi:10.1109/ICCET.2009.73.

[14]\qquad Zhao Z., Zhu J., Xia B., Multi-fractal fluctuation features of
thermal power coal price in China. Energy, 117(2016),10--8.
doi:10.1016/j.energy.2016.10.081.

[15]\qquad Salv\'{o} G., Piacquadio M.N., Multifractal analysis of
electricity demand as a tool for spatial forecasting. Energy Sustain Dev,
38(2017), 67--76. doi:10.1016/j.esd.2017.02.005.

[16]\qquad Zhai M-Y. A new method for short-term load forecasting based on
fractal interpretation and wavelet analysis. Int J Electr Power Energy Syst,
69(2015), 241--245. doi:10.1016/j.ijepes.2014.12.087.

[17]\qquad Oszutowska D., Purczynski J., Estimation of the Fractal Dimension
using Tiled Triangular Prism Method for Biological Non-Rectangular Objects,
Przeglad Elektrotechniczny,88(2012), nr 10b, 261--263.

[18]\qquad Bekasiewicz A., Kurgan P., Duraj P., Compact antenna array
comprising fractal-shaped microstrip radiators, Przeglad Elektrotechniczny,
87(2011), nr 10, 98--100.

[19] Katok A., Hasselblatt B., Introduction to the modern theory of
dynamical systems. New York, New York, USA, (1996). doi:10.1137/1038033

[20]\qquad Borjon J., Caos, orden y desorden en el sistema monetario y
financiero internacional. ed. 1, (2002).

\section{Bibliography of authors}

\includegraphics{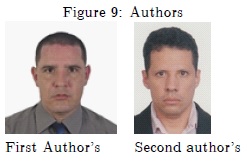}

\begin{figure*}[h]\centering%
\caption{Authors}%
\label{Fig9}%
\begin{tabular}{ll}
&  \\ 
First Author's & Second author's%
\end{tabular}%
\end{figure*}%

Firts Author's: H\'{e}ctor A. Tabares-Ospina: received his Bs. degree in
Electrical Engineering in 1997 and his Master in Systems Engineering in 2005
from Universidad Nacional de Colombia. He is now studing doctoral studies.
He is an Assistant Professor of Instituci\'{o}n Universitaria Pascual Bravo.
His research interests include: Fractal geometry, artificial intelligence,
operation and control of power systems; and smart grids. He is a Junior
Researcher in Colciencias and member of the Research Group - GIIEN, at
Instituci\'{o}n Universitaria Pascual Bravo.
https://orcid.org/0000-0003-2841-6262\\*[0pt]

Second Author's:John E. Candelo-Becerra: received his Bs. degree in
Electrical Engineering in 2002 and his PhD in Engineering with emphasis in
Electrical Engineering in 2009 from Universidad del Valle, Cali - Colombia.
His employment experiences include the Empresa de Energ\'{\i}a del Pac\'{\i}%
fico EPSA, Universidad del Norte, and Universidad Nacional de Colombia -
Sede Medell\'{\i}n. He is now an Assistant Professor of the Universidad
Nacional de Colombia - Sede Medell\'{\i}n, Colombia. His research interests
include: engineering education; planning, operation and control of power
systems; artificial intelligence; and smart grids. He is a Senior Researcher
in Colciencias and member of the Applied Technologies Research Group - GITA,
at the Universidad Nacional de Colombia.
https://orcid.org/0000-0002-9784-9494.

\end{document}